\documentclass{PoS}
\usepackage{amsmath}
\usepackage{url}

\title{Current status of Two-Higgs-Doublet models with a softly broken $\mathbb{Z}_2$ symmetry}

\ShortTitle{Current status of 2HDMs}

\author{\speaker{Otto Eberhardt}\\
        Instituto de F\'{i}sica Corpuscular, Universitat de Val\`{e}ncia -- Consejo Superior de
        Investigaciones Cient\'{i}ficas, Apt. Correus 22085, E-46071 Val\`{e}ncia, Spain\\
        E-mail: \email{otto.eberhardt@ific.uv.es}}

\abstract{One of the most popular extensions of the Standard Model is the Two-Higgs-Doublet model (2HDM),
in which a second Higgs doublet is added to the conventional Standard Model particle content.
2HDM's with a softly broken $\mathbb{Z}_2$ symmetry avoid flavour-changing neutral Higgs currents at tree-level.
They also comprise the Higgs sector of the Minimal Supersymmetric Standard Model.
Current Higgs observables put strong constraints on these models.
We present global fits to these data and combine them with information from theoretical bounds,
electroweak precision observables and the most important flavour constraints, using the open-source package HEPfit.
The resulting limits on the 2HDM parameters like mixing angles,
the heavy Higgs masses as well as the allowed Higgs mass splittings will be discussed.
}

\FullConference{XXXIX International Conference for High-Energy Physics\\
		 4-11 July 2018\\
		 Seoul, Republic of Korea}

\begin{document}

\section{Introduction}

The Two-Higgs Doublet model \cite{Lee:1973iz,Branco:2011iw} adds a second scalar $SU(2)$ doublet to the Standard Model (SM) fields. In its most general formulation, it can offer a solution to several unsolved questions in flavour physics. However, recent LHC data show that the Higgs particle found at $125$ GeV behaves much like the SM Higgs; at current experimental precision there are no signs for flavour-changing or CP violating Higgs interactions.
What I will denote as 2HDM in the following is a Two-Higgs doublet model with a $\mathbb{Z}_2$ symmetry, which is only softly broken, and the absence of CP violation in the Higgs sector, yielding an SM-like Higgs as the lightest massive particle of the scalar sector.
While opening a window to non-minimal Higgs sectors (like the one of the Minimal Supersymmetric Standard Model) and stabilising the scalar potential with respect to the SM one, large parts of this model are strongly constrained by various experimental and theoretical constraints. Using the {\tt HEPfit} package \cite{hepfit}, I apply a comprehensive statistical analysis in order to identify the allowed parameter values and to make quantitative statements about the excluded parameter regions.

\section{The 2HDM}

The introduction of a second Higgs doublet increases the number of independent Lorentz invariant terms of the scalar potential from two in the SM to eight in the 2HDM:\\[-30pt]

\begin{align*}
V_H^{\text{\tiny{2HDM}}}=
& \; m^2_{11} \Phi_1^\dagger \Phi_1
+m^2_{22}\Phi_2^\dagger \Phi_2^{\phantom{\dagger}}
-m_{12}^2 \left( \Phi_1^\dagger \Phi_2^{\phantom{\dagger}}
+ \Phi_2^\dagger \Phi_1^{\phantom{\dagger}} \right) +\frac{\lambda_1}{2} \left( \Phi^\dagger _1\Phi^{\phantom{\dagger}}_1 \right) ^2
+\frac{\lambda_2}{2} \left( \Phi_2^{\dagger}\Phi_2^{\phantom{\dagger}} \right) ^2\\
& \; +\lambda_3 \left( \Phi_1^{\dagger}\Phi_1^{\phantom{\dagger}}\right)  \left( \Phi_2^{\dagger}\Phi_2^{\phantom{\dagger}}\right) +\lambda_4 \left( \Phi_1^{\dagger}\Phi_2^{\phantom{\dagger}}\right)  \left( \Phi_2^{\dagger}\Phi_1^{\phantom{\dagger}}\right) 
+\frac{\lambda_5}{2}\left[ \left( \Phi_1^{\dagger}\Phi_2^{\phantom{\dagger}}\right) ^2
+\left( \Phi_2^{\dagger}\Phi_1^{\phantom{\dagger}}\right) ^2 \right] \\[-20pt]
\end{align*}

The $\mathbb{Z}_2$ symmetry, under which one of the doublets is odd, has been applied in order to avoid flavour-changing neutral Higgs interactions at tree-level; it is only softly broken by $m_{12}^2$. A consequence of this symmetry is that $t$ quarks only couple to $\Phi_2$ and $b$ and $\tau$ only to $\Phi_2$ ($\Phi_1$) in the so-called 2HDM of type I (II).
All parameters in $V_H^{\text{\tiny{2HDM}}}$ are assumed to be real.
If we rotate the generic Higgs fields to the mass eigenstates, we obtain three neutral Higgs particles $h$, $H$ and $A$ and two charged mass eigenstates $H^\pm$, of which we identify $h$ with the SM-like Higgs found at the LHC, while the other particles are assumed to be heavier.
In the global Bayesian fits with the {\tt HEPfit} interface to BAT \cite{Caldwell:2008fw}, I apply the following priors to the parameters: The ratio between the vacuum expectation values of $\Phi_2$ and $\Phi_1$, called $\tan \beta$, is varied between $0.1$ and $50$; $\beta\!-\!\alpha$, which is the difference between the diagonalisation angles of the CP odd and CP even fields, floats between $0$ and $\pi$ (with $\pi/2$ reproducing an $h$ with SM-like couplings \cite{Gunion:2002zf}); the masses $m_H$, $m_A$, $m_{H^+}$ are assumed to be between $130$ and $1600$ GeV; and $|m_{12}^2|$ cannot exceed $1.6$ TeV.

\section{Constraints}
\label{sec:Constraints}

The following constraints are used in the global 2HDM fits:
The scalar potential $V_H^{\text{\tiny{2HDM}}}$ is required to be bounded from below \cite{Deshpande:1977rw} and the electroweak vacuum is assumed to be its global minimum \cite{Barroso:2013awa}. The probability of a transition of two scalars into two scalars should not exceed $1$ \cite{Ginzburg:2005dt}. This is guaranteed at the one-loop level together with the assumption of convergence of the perturbation series, i.e.~enforcing the next-to-leading order contribution to be smaller in magnitude than the tree-level expression \cite{Grinstein:2015rtl,Cacchio:2016qyh}. I will refer to the previously mentioned constraints as theoretical constraints in the following.
As experimental constraints I use the $h$ signal strengths and the LHC searches for neutral and singly charged scalars. Moreover, electroweak precision observables are applied using the $STU$ framework, and from the flavour sector the measurements of $b\to s \gamma$ decays and $B_s$ meson mixing are forced to stay within their experimental uncertainties. I refer to the references \cite{Cacchio:2016qyh,Chowdhury:2017aav} for details about the implementation of all of the constraints.

\section{Fit results}

\begin{figure}
\begin{center}
\begin{picture}(350,200)(0,0)
\put(0,-10){\includegraphics[width=350pt]{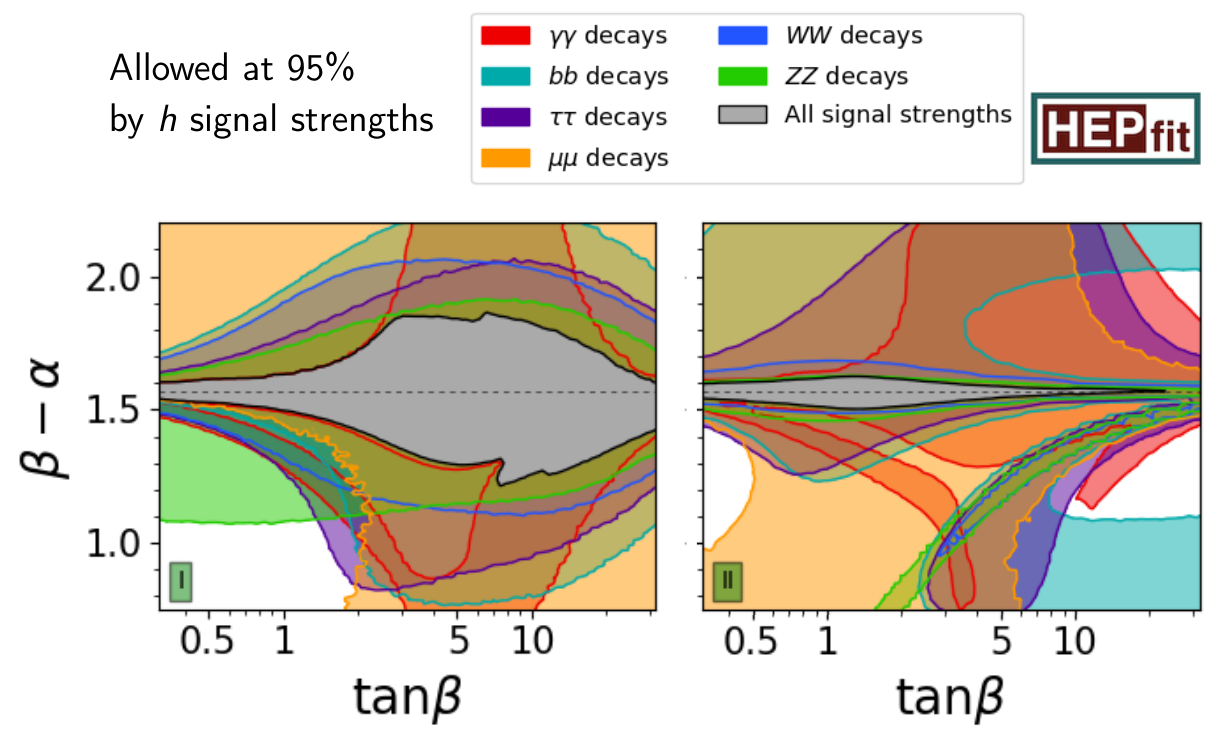}}
\end{picture}
\end{center}
\caption{Breakdown of the $95$\% regions allowed by $h$ signal strengths into decay channels in the $\beta\!-\!\alpha$ vs.~$\tan \beta$ planes. The left (right) panel shows the fit in the 2HDM of type I (II). For the coloured contours I refer to the legend; the black-bordered grey region ``survives'' if one uses all signal strengths in the fit. The dotted line denotes the limit in which the $h$ couplings become SM-like.}
\label{fig:1}
\end{figure}

In Fig.~\ref{fig:1} I show the impact of only the $h$ signal strengths on the $\beta\!-\!\alpha$ vs.~$\tan \beta$ plane of the 2HDM of type I and II. The coloured contours delimit the regions which are allowed with a probability of $95\%$, if one only uses the experimental information of a single $h$ decay channel. The grey regions, which roughly correspond to the superpositions of the shaded contours, show the $95\%$ allowed values for $\beta\!-\!\alpha$ and $\tan \beta$, if all $h$ signal strengths are used as fit constraints. The deviation of $\beta\!-\!\alpha$ from the SM limiting case cannot exceed 0.32 (0.06) in the type I (type II) fit.

\begin{figure}
\begin{center}
\begin{picture}(350,150)(0,0)
\put(0,0){\includegraphics[width=365pt]{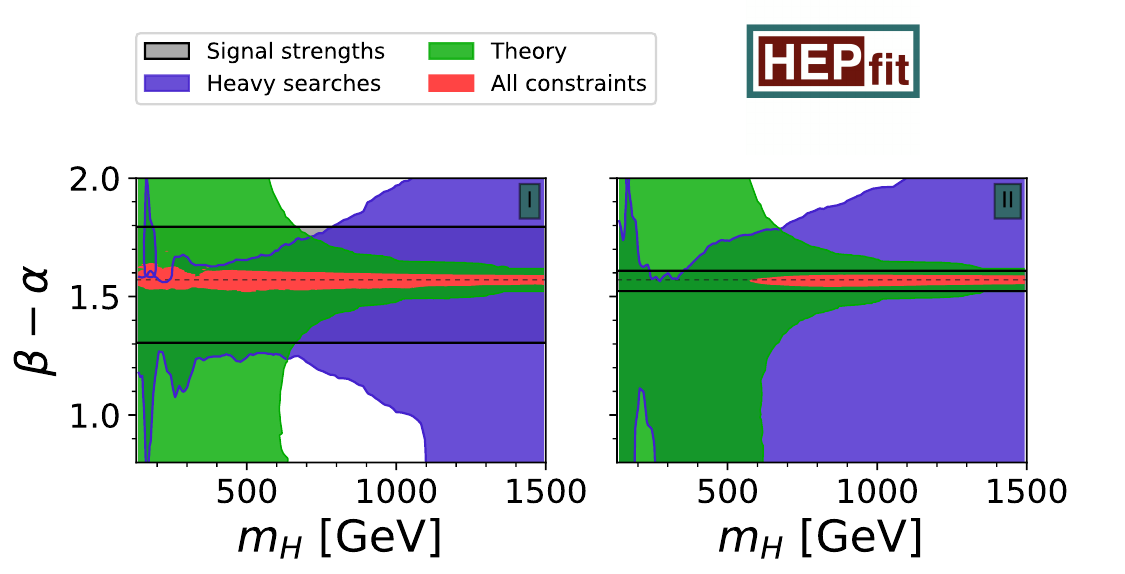}}
\end{picture}
\begin{picture}(350,300)(0,0)
\put(0,0){\includegraphics[width=350pt]{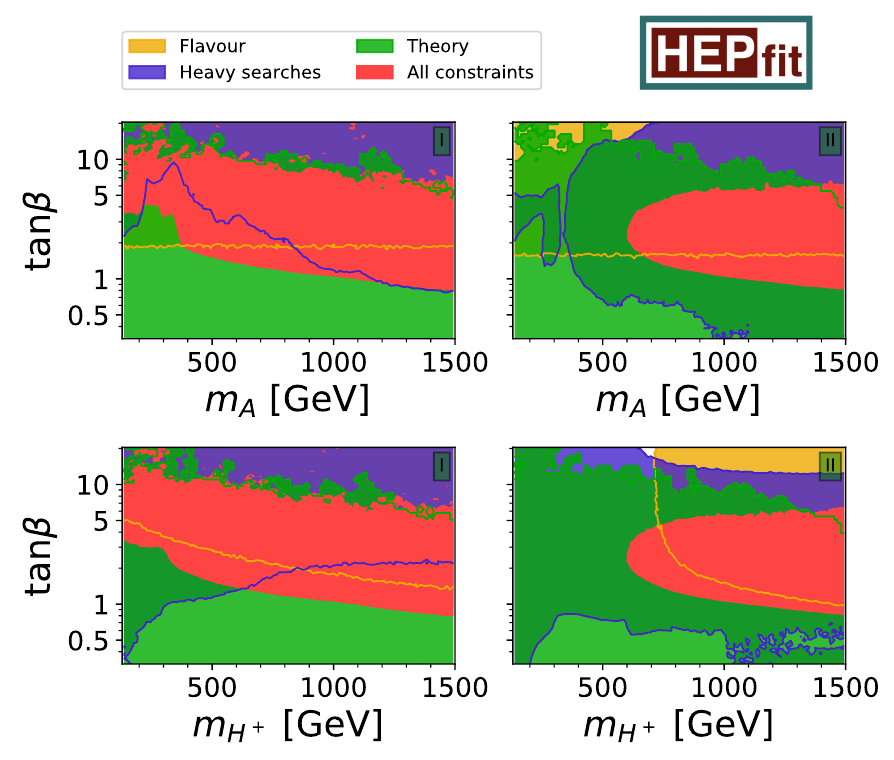}}
\end{picture}
\end{center}
\caption{Impact of the various sets of constraints on the $\beta\!-\!\alpha$ vs.~$m_H$ (upper row), the $\tan \beta$ vs.~$m_{A/H^+}$ (lower rows) planes in the 2HDM of type I (left) and type II (right). The individually applied sets of constraints are the $h$ signal strengths (grey contours), the theoretical bounds (in green), the searches for heavy Higgs particles (in blue) and the flavour observables (in yellow). Their combination is given by the red contours. All shaded regions are allowed with a probability of $95\%$.}
\label{fig:2}
\end{figure}

The results of the fits with only certain sets of the constraints mentioned in Sec.~\ref{sec:Constraints} as well as the simultaneous fit to all constraints in the $\beta\!-\!\alpha$ vs.~$m_H$ and the $\tan \beta$ vs.~$m_{A/H^+}$ planes of the 2HDM of type I and II can be found in Fig.~\ref{fig:2}. 
In both types, the difference between the mixing angles $\beta$ and $\alpha$ can deviate by at most 0.03 from the alignment limit $\pi/2$ with $95\%$ probability.
In type I scenarios with relatively low $\tan \beta$ and light $m_{A/H^+}$ can be excluded at $95\%$. In the 2HDM of type II, one can observe a lower bound on the heavy Higgs masses of around $600$ GeV, and $|m_{12}^2|$ cannot exceed $(280 \;\text{GeV})^2$. Thus the possibility of an unbroken $\mathbb{Z}_2$ symmetry is excluded. These limits stem from the lower bound on the charged Higgs mass by $b\to s \gamma$ decays and the fact that due to unitarity and electroweak precision data the mass differences between the heavy Higgs particles cannot be larger than $130$ GeV. Also in the 2HDM of type I, we extract upper limits on these mass splittings; they can maximally be as large as $200$ GeV. All decays of $H$ into another heavy 2HDM scalar (like $H\to AA$ or $H\to H^+W^-$) can be excluded in type I and II.
Finally, I analyse the total decay widths of the heavy Higgs bosons, and find that they cannot be larger than $7\%$ ($5\%$) in type I (II) with a probability of $95\%$.

\section{Summary}

I have presented the results of global fits to the 2HDM of type I and II obtained with the open-source package {\tt HEPfit}. Some parameters and parameter combinations are strongly constrained by the existing theoretical and experimental constraints.

\acknowledgments

I thank Debtosh Chowdhury for helpful discussions and proofreading.
This work was supported by Grants No.~FPA2014- 53631-C2-1-P, FPA2017-84445-P and SEV-2014-0398 (AEI/ERDF, EU).

\bibliographystyle{JHEP}
\bibliography{Current_status_of_2HDMs}

\providecommand{\href}[2]{#2}\begingroup\raggedright\begin{thebibliography}{10}

\bibitem{Lee:1973iz}
T.~Lee, \emph{{A Theory of Spontaneous T Violation}},
  \href{https://doi.org/10.1103/PhysRevD.8.1226}{\emph{Phys.Rev.} {\bfseries
  D8} (1973) 1226--1239}.

\bibitem{Branco:2011iw}
G.~Branco, P.~Ferreira, L.~Lavoura, M.~Rebelo, M.~Sher et~al., \emph{{Theory
  and phenomenology of two-Higgs-doublet models}},
  \href{https://doi.org/10.1016/j.physrep.2012.02.002}{\emph{Phys.Rept.}
  {\bfseries 516} (2012) 1--102},
  [\href{https://arxiv.org/abs/1106.0034}{{\ttfamily 1106.0034}}].

\bibitem{hepfit}
{\scshape HEPfit} collaboration, ``{\tt HEPfit}: a code for the combination of
  indirect and direct constraints on high energy physics models.'' in
  preparation. \url{http://hepfit.roma1.infn.it}.

\bibitem{Caldwell:2008fw}
A.~Caldwell, D.~Kollar and K.~Kroninger, \emph{{BAT: The Bayesian Analysis
  Toolkit}}, \href{https://doi.org/10.1016/j.cpc.2009.06.026}{\emph{Comput.
  Phys. Commun.} {\bfseries 180} (2009) 2197--2209},
  [\href{https://arxiv.org/abs/0808.2552}{{\ttfamily 0808.2552}}].

\bibitem{Gunion:2002zf}
J.~F. Gunion and H.~E. Haber, \emph{{The CP conserving two Higgs doublet model:
  The Approach to the decoupling limit}},
  \href{https://doi.org/10.1103/PhysRevD.67.075019}{\emph{Phys. Rev.}
  {\bfseries D67} (2003) 075019},
  [\href{https://arxiv.org/abs/hep-ph/0207010}{{\ttfamily hep-ph/0207010}}].

\bibitem{Deshpande:1977rw}
N.~G. Deshpande and E.~Ma, \emph{{Pattern of Symmetry Breaking with Two Higgs
  Doublets}}, \href{https://doi.org/10.1103/PhysRevD.18.2574}{\emph{Phys. Rev.}
  {\bfseries D18} (1978) 2574}.

\bibitem{Barroso:2013awa}
A.~Barroso, P.~M. Ferreira, I.~P. Ivanov and R.~Santos, \emph{{Metastability
  bounds on the two Higgs doublet model}},
  \href{https://doi.org/10.1007/JHEP06(2013)045}{\emph{JHEP} {\bfseries 06}
  (2013) 045}, [\href{https://arxiv.org/abs/1303.5098}{{\ttfamily 1303.5098}}].

\bibitem{Ginzburg:2005dt}
I.~F. Ginzburg and I.~P. Ivanov, \emph{{Tree-level unitarity constraints in the
  most general 2HDM}},
  \href{https://doi.org/10.1103/PhysRevD.72.115010}{\emph{Phys. Rev.}
  {\bfseries D72} (2005) 115010},
  [\href{https://arxiv.org/abs/hep-ph/0508020}{{\ttfamily hep-ph/0508020}}].

\bibitem{Grinstein:2015rtl}
B.~Grinstein, C.~W. Murphy and P.~Uttayarat, \emph{{One-loop corrections to the
  perturbative unitarity bounds in the CP-conserving two-Higgs doublet model
  with a softly broken $ {\mathrm{\mathbb{Z}}}_2 $ symmetry}},
  \href{https://doi.org/10.1007/JHEP06(2016)070}{\emph{JHEP} {\bfseries 06}
  (2016) 070}, [\href{https://arxiv.org/abs/1512.04567}{{\ttfamily
  1512.04567}}].

\bibitem{Cacchio:2016qyh}
V.~Cacchio, D.~Chowdhury, O.~Eberhardt and C.~W. Murphy, \emph{{Next-to-leading
  order unitarity fits in Two-Higgs-Doublet models with soft $\mathbb{Z}_2$
  breaking}}, \href{https://doi.org/10.1007/JHEP11(2016)026}{\emph{JHEP}
  {\bfseries 11} (2016) 026},
  [\href{https://arxiv.org/abs/1609.01290}{{\ttfamily 1609.01290}}].

\bibitem{Chowdhury:2017aav}
D.~Chowdhury and O.~Eberhardt, \emph{{Update of Global Two-Higgs-Doublet Model
  Fits}}, \href{https://doi.org/10.1007/JHEP05(2018)161}{\emph{JHEP} {\bfseries
  05} (2018) 161}, [\href{https://arxiv.org/abs/1711.02095}{{\ttfamily
  1711.02095}}].

\end{thebibliography}\endgroup

\end{document}